\documentclass[preprint,superscriptaddress]{revtex4}
\usepackage{graphicx}
\usepackage{qcircuit}

\newcommand{\ket}[1] {\left| #1 \right\rangle}

\begin{document}

\title{Quantum circuit synthesis via a random combinatorial search}

\author{Sahel Ashhab}
\affiliation{Advanced ICT Research Institute, National Institute of Information and Communications Technology (NICT), 4-2-1, Nukui-Kitamachi, Koganei, Tokyo 184-8795, Japan}

\author{Fumiki Yoshihara}
\affiliation{Advanced ICT Research Institute, National Institute of Information and Communications Technology (NICT), 4-2-1, Nukui-Kitamachi, Koganei, Tokyo 184-8795, Japan}
\affiliation{Department of Physics, Tokyo University of Science, 1-3 Kagurazaka, Shinjuku-ku, Tokyo 162-8601, Japan}

\author{Miwako Tsuji}
\affiliation{RIKEN Center for Computational Science, Kobe, Hyogo, 650-0047, Japan}

\author{Mitsuhisa Sato}
\affiliation{RIKEN Center for Computational Science, Kobe, Hyogo, 650-0047, Japan}

\author{Kouichi Semba}
\affiliation{Advanced ICT Research Institute, National Institute of Information and Communications Technology (NICT), 4-2-1, Nukui-Kitamachi, Koganei, Tokyo 184-8795, Japan}
\affiliation{Institute for Photon Science and Technology, The University of Tokyo, 7-3-1 Hongo, Bunkyo-ku, Tokyo 113-0033, Japan}

\begin{abstract}
We use a random search technique to find quantum gate sequences that implement perfect quantum state preparation or unitary operator synthesis with arbitrary targets. This approach is based on the recent discovery that there is a large multiplicity of quantum circuits that achieve unit fidelity in performing a given target operation, even at the minimum number of single-qubit and two-qubit gates needed to achieve unit fidelity. We show that the fraction of perfect-fidelity quantum circuits increases rapidly as soon as the circuit size exceeds the minimum circuit size required for achieving unit fidelity. This result implies that near-optimal quantum circuits for a variety of quantum information processing tasks can be identified relatively easily by trying only a few randomly chosen quantum circuits and optimizing their parameters. In addition to analyzing the case where the CNOT gate is the elementary two-qubit gate, we consider the possibility of using alternative two-qubit gates. In particular, we analyze the case where the two-qubit gate is the B gate, which is known to reduce the minimum quantum circuit size for two-qubit operations. We apply the random search method to the problem of decomposing the 4-qubit Toffoli gate and find a 15 CNOT-gate decomposition.
\end{abstract}

\maketitle

\section{Introduction}
\label{Sec:Introduction}

One of the important steps in implementing a quantum algorithm or performing a multi-qubit operation on a quantum computing device is designing a sequence of elementary quantum gates to perform the desired multi-qubit transformation \cite{Barenco,Knill,ShendeUnitary,Bergholm,Mottonen,AshhabQuantumCircuits,Nielsen} or prepare the desired multi-qubit state \cite{Grover,Kraus,Kaye,ShendeStatePrep,Soklakov}. This step can be thought of as compiling the quantum code from a high-level language to the low level of single- and two-qubit gates. The elementary quantum gate sequence is often called the quantum circuit.

Different technologies are currently being investigated for future quantum computing devices \cite{Ladd,Buluta}. The specific technology used in an experimental setup often dictates how the qubits are addressed. For example, the most natural way to manipulate trapped ions \cite{Sorensen,Martinez,GoubaultDeBrugiere2019} is different from the most natural way to manipulate superconducting qubits \cite{Paraoanu,Rigetti,DeGroot}, and each technology has its own set of constraints on the allowed qubit control tools. Physical constraints on control signals can lead to technology-specific preferences in the elementary gate set used in the compilation step, i.e.~in quantum circuit design \cite{AshhabSpeedLimits2012}. While these experimental considerations must be considered in practical settings, past theoretical studies on quantum circuit synthesis have generally focused on elementary gate sets based on the controlled-NOT (CNOT) gate as the two-qubit entangling gate, combined with the set of all possible single-qubit unitary operators. We follow this convention and focus on the CNOT gate. It should be noted that the CNOT gate is a natural gate for numerous superconducting device architectures in use today \cite{Rigetti,DeGroot}.

An important question in the study of quantum circuit synthesis is quantum circuit size or complexity, which is quantified by the number of entangling gates $N$. In particular, it is highly desirable to find quantum circuits that have the minimum number of entangling gates needed for a perfect implementation of arbitrary multi-qubit operations, including both state preparation and unitary operator synthesis. For this purpose, it is of great value to know in advance the minimum required circuit size. This question has been studied in numerous previous studies, typically by either devising recipes for quantum circuit synthesis \cite{Znidaric,Plesch,Vidal,Vatan,Vartiainen,GoubaultDeBrugiere2020,Rakyta} or by using mathematical arguments to derive lower and upper bounds for the required circuit size \cite{ShendeUnitary,ShendeStatePrep}. In Ref.~\cite{AshhabQuantumCircuits}, we investigated this question by numerically analyzing all the possible gate sequences up to a certain size for few-qubit tasks. Our results provided definite values for the minimum number of gates needed for various tasks, hence serving as tests and benchmarks to assess previously calculated lower bounds and recipe-based circuit sizes. There have been a few other recent studies on computer-based quantum circuit synthesis \cite{Camps,ZhangAI2021,Shirakawa,ZhangAI2022,Wang2023,Szasz,Weiden,Rosenhahn,Daimon,Wang2024}.

One of the remarkable results that we found in Ref.~\cite{AshhabQuantumCircuits} is the large multiplicity of quantum circuits that can be used to perform the same task. In particular, even at the minimum number of CNOT gates needed for perfect state preparation of an arbitrary four-qubit state or unitary operator synthesis of an arbitrary three-qubit operator, we found that 20\% of all possible CNOT gate configurations allow performing the desired task perfectly, i.e.~with unit fidelity. This result naturally points to a possibly very efficient method to find optimal or near-optimal quantum circuits for various tasks. Although the enormous number of different gate configurations limited our previous study to small numbers of qubits, if a finite fraction of all possible configurations give unit fidelity, it suffices to analyze a few randomly chosen quantum circuits. One will then, with a high probability, find at least one quantum circuit that performs the desired task perfectly.

In this work, we demonstrate this probabilistic approach to quantum circuit synthesis and treat cases that are intractable using the exhaustive search method. Our results indicate that the large multiplicity of unit-fidelity quantum circuits continues to hold for larger numbers of qubits than those investigated previously. Furthermore, we show that finding a unit-fidelity quantum circuit via a random search becomes increasingly easy if we increase the number of gates only slightly above the minimum required circuit size.

Since one of our main goals is reducing the gate count, we also consider the so-called $B$ gate \cite{ZhangBGate}. The reason behind this choice is that arbitrary two-qubit operations require shorter gate sequences if one uses the $B$ gate instead of the CNOT gate \cite{ZhangBGate}. We show that the $B$ gate allows constructing shorter gate sequences than the CNOT gate for larger numbers of qubits. This result can be understood intuitively based on commutation relations and simple parameter-counting arguments. We emphasize, however, that experimental difficulties in implementing the $B$ gate could outweigh any benefits that it provides, as we shall discuss below.

The remainder of this paper is organized as follows: in Sec.~\ref{Sec:Background}, we provide some background information about quantum circuit synthesis and our computational approach. In Sec.~\ref{Sec:Results}, we present the results of our numerical calculations and discuss their implications for practical quantum circuit synthesis. In Sec.~\ref{Sec:Conclusion}, we give some concluding remarks.

\section{Background}
\label{Sec:Background}

\subsection{Theoretical background}

As mentioned in Sec.~\ref{Sec:Introduction}, implementing a multi-qubit operation on a quantum computer is typically achieved by finding a decomposition of the desired operation into a sequence of one- and two-qubit gates and then implementing that sequence. The CNOT gate, which is readily implementable on several quantum computing platforms, is the standard two-qubit gate used in most theoretical studies on the subject of quantum circuit synthesis.

Synthesizing a quantum circuit that implements an arbitrary $n$-qubit operation is analogous to reaching an arbitrary target point in the space of quantum operations \cite{AshhabQuantumCircuits}. Each added quantum gate adds adjustable control parameters that allow us to access more regions in the space of quantum operations. Lower bounds for the number of CNOT gates needed to perform an arbitrary $n$-qubit operation perfectly can be obtained by counting the free parameters in an arbitrary target and the number of adjustable parameters in the quantum circuit. The quantum circuit parameters must be at least equal to the number of free target parameters to be able to cover the space of all targets. This parameter counting calculation gives the lower bounds
\begin{equation}
N_{\rm LB, SP, CNOT} = \left\lceil \frac{2^{n}-1-n}{2} \right\rceil
\label{Eq:LowerBoundStatePrep}
\end{equation}
for state preparation and
\begin{equation}
N_{\rm LB, U, CNOT} = \left\lceil \frac{4^n-1-3n}{4} \right\rceil
\label{Eq:LowerBoundUnitary}
\end{equation}
for unitary operator synthesis. Here the function $\lceil x \rceil$ is the ceiling function, i.e.~the smallest integer larger than or equal to $x$.

Another important quantity in the present context is the number of different configurations for a single CNOT gate, i.e.~the number of distinct ways that we can select a pair out of $n$ qubits. Since reversing the control and target qubits in a CNOT gate gives the same CNOT gate, up to single-qubit rotations, the number of distinct CNOT gate configurations for a single gate is $n(n-1)/2$. As a result, the total number of CNOT gate configurations for a quantum circuit of size $N$ is
\begin{equation}
N_{\rm config} = \left( \frac{n(n-1)}{2} \right)^N.
\label{Eq:Nconfig}
\end{equation}

In Ref.~\cite{AshhabQuantumCircuits} we performed calculations in which we evaluated the maximum achievable fidelity for all possible CNOT gate configurations for $n$ qubits, with $n\leq 4$ for state preparation and $n\leq 3$ for unitary operator synthesis. In the latter case, the computation time was on the order of a few years if the computations were run on a single core of a present-day computer. We now consider what would happen if we try to perform the same calculations for a larger number of qubits. Table \ref{Table:ResourceEstimates} shows the number of CNOT gate configurations for different system sizes. If we go from $n=3$ to $n=4$ in the case of unitary operator synthesis, a number of factors change and dramatically increase the computation time. The Hilbert space size doubles, and the number of CNOT gates needed to achieve unit fidelity roughly quadruples. More importantly, the number of gate configurations $N_{\rm config}$ increases by a factor of $\sim 10^{41}$, which is a result of the super-exponential scaling described by Eqs.~(\ref{Eq:LowerBoundStatePrep}-\ref{Eq:Nconfig}). It is therefore impossible to extend the calculations of Ref.~\cite{AshhabQuantumCircuits} to larger systems.

\begin{center}
\begin{table}
\scriptsize
\begin{tabular}{ || c || c | c | c || c | c | c || }
\hline
& \multicolumn{3}{|c||}{State preparation} & \multicolumn{3}{|c||}{Unitary operator synthesis} \\
\hline
No. of qubits & $N_{\rm CNOT}$ & \, \, \, $N_{\rm config}$ \, \, \, & Comp. time per config. & $N_{\rm CNOT}$ & \, \, \, $N_{\rm config}$ \, \, \, & Comp. time per config. \\
\hline
2 & 1 & 1 & $\sim$ 0.1 sec. & 3 & 1 & $\sim$ 1 sec. \\
\hline
3 & 2 & 9 & $\sim$ 2 sec. & 14 & $4.8 \times 10^6$ & $\sim$ 10 min. \\
\hline
4 & 6 & $4.7 \times 10^4$ & $\sim$ 1 min. & 61 & $\sim 10^{47}$ & $\sim$ 1 hour \\
\hline
5 & 13 & $\sim 10^{13}$ & $\sim$ 5 min. & 252 & $\sim 10^{252}$ & $\sim$ 10 hours \\
\hline
6 & 29 & $\sim 10^{34}$ & $\sim$ 1 hour & & & \\
\hline
7 & 60 & $\sim 10^{79}$ & $\sim$ 10 hours & & & \\
\hline
8 & 124 & $\sim 10^{179}$ & $\sim$ 20 hours & & & \\
\hline
\end{tabular}
\caption{Estimates for the quantum circuit size $N_{\rm CNOT}$, the total number of possible gate configurations $N_{\rm config}$ and computation time per configuration as functions of system size, i.e.~the number of qubits $n$. The number $N_{\rm CNOT}$ for each case is calculated from the appropriate lower bound formula, i.e.~Eq.~(\ref{Eq:LowerBoundStatePrep}) or (\ref{Eq:LowerBoundUnitary}). The number $N_{\rm config}$ is calculated from Eq.~(\ref{Eq:Nconfig}), with $N$ given by the appropriate value of $N_{\rm CNOT}$. The computation times are specific to our computational environment: each calculation was run on a single core up to $n=4$ and on about 10 cores for $n\geq 5$. The calculations for $n\leq 7$ were  performed on workstations, while those for $n=8$ were performed on the Fugaku supercomputer. By far, the quantity that grows the fastest with increasing $n$ is $N_{\rm config}$.}
\label{Table:ResourceEstimates}
\end{table}
\end{center}

\subsection{Computational approach}

Since it is practically impossible to analyze all the gate configurations for the system sizes that we investigate in this work, we use a probabilistic approach. We analyze a number ($N_{\rm samples}$) of randomly generated quantum circuits, each defined by the CNOT gate configuration. The number $N_{\rm samples}$ is a small fraction of the total number of possible gate configurations. Importantly, it does not scale with system size, and hence we avoid the super-exponential scaling mentioned in the previous subsection. As a general rule, for each set of parameters $n$ and $N$, we set $N_{\rm samples}=100$. The reason why we choose this number is that it is sufficiently large to allow us to perform statistical analysis for the purposes of the present study.

The target operation is generated randomly, as described in Ref.~\cite{AshhabQuantumCircuits}. This  target can be thought of as a random point in the space of operations or a random sample from a Haar-measure uniform distribution of quantum states or unitary operators.

Once we have specified the target, we generate gate sequences with the aim of reaching the target from a fixed starting point. Each gate sequence is defined by a CNOT gate configuration along with the appropriate single-qubit rotations surrounding the CNOT gates. For each such configuration, the single-qubit rotation parameters serve as adjustable parameters that are chosen based on the target. They are optimized using a modified version of the gradient ascent pulse engineering (GRAPE) algorithm \cite{AshhabQuantumCircuits,Khaneja}. In the numerical optimization calculations, the fidelity $F$ is used as the quantity to be maximized. For state preparation,
\begin{equation}
F = {\rm Tr} \left\{ \rho_F U(T) \rho_0 U^{\dagger}(T) \right\},
\end{equation}
where $\rho_0$ is the initial density matrix (with all the qubits in the state $\ket{0}$), $\rho_F$ is the target density matrix, and $U(T)$ is the evolution operator generated by the quantum circuit, whose parameters are being updated by the optimization algorithm. For unitary operator synthesis,
\begin{equation}
F = \left| \frac{ {\rm Tr} \left\{ U_F^{\dagger} U(T) \right\} }{2^n} \right|^2,
\end{equation}
where $U_F$ is the target unitary matrix.

In each iteration of the GRAPE algorithm, the derivatives of the fidelity with respect to all possible variations in all single-qubit rotations are calculated, which results in a gradient of the fidelity with respect to the rotation parameters. The rotations are then updated such that the fidelity increases steadily as the algorithm progresses. The optimization procedure is repeated for up to $10^5$ iterations. The algorithm is terminated early if the fidelity exceeds $1-10^{-12}$ or grows by less than $10^{-12}$ in $10^3$ iterations.

If the optimization gives $1-F<10^{-8}$ for a given quantum circuit, we identify the quantum circuit as corresponding to $F=1$, i.e.~a quantum circuit that implements the target perfectly. The error threshold, i.e.~the number $10^{-8}$, is chosen because it is sufficiently small that any deviation from $F=1$ below this threshold can be attributed to numerical errors. Besides, from the perspective of practical implementations, an error of $10^{-8}$ is almost certain to be much smaller than other realistic errors in the experimental setup.

Since we would like to establish a probabilistic approach to quantum circuit synthesis, we use a probability-based interpretation of our results. The probability ($p$) that a random quantum circuit will give $F=1$ in a given setting is straightforwardly inferred from the fraction ($p_{\rm num}$) of the circuits that give $F=1$ in our simulations, i.e.~$p=p_{\rm num}$. To determine the uncertainty in the calculated value of $p$, we need to evaluate the statistical variations in the results of the numerical calculations. We can do so by calculating the standard deviation $\sigma$ in the data. Since each randomly generated circuit is classified as either having $F=1$ or not, we need to use the formula for a binomial distribution. A quick and simple estimate for $\sigma$ can be obtained by treating the value of $p_{\rm num}$ obtained from the simulations as the actual probability $p$ in the binomial distribution. We then substitute this value in the formula for the binomial distribution standard deviation,
\begin{equation}
\sigma = \sqrt{\frac{p(1-p)}{N_{\rm samples}}}.
\label{Eq:BinomialSTD}
\end{equation}
This simple formula gives accurate results for large values of $N_{\rm samples}$ and $p$ not too close to the end points $p=0$ or $p=1$. Since some of our data sets give $p_{\rm num}=0$, we avoid the simple calculation above and instead perform a Bayesian probability calculation to determine the size of the error bars and evaluate the uncertainty in $p$. We divide the range of $p$ values, i.e.~the range [0, 1], into a large number of bins, which results in a statistical ensemble that we call ``Bins'', with each bin characterized by a $p$ value. For each data set, with a given fraction $p_{\rm num}$ of perfect-fidelity quantum circuits, we calculate the probability of obtaining $p_{\rm num}$ for each value of $p$ in the ensemble. We refer to this conditional probability as $P(p_{\rm num}|p)$. Using Bayes' rule, the probability distribution for $p$ values for a numerically obtained value of $p_{\rm num}$ is given by
\begin{equation}
P(p|p_{\rm num}) = \frac{P(p_{\rm num}|p)}{\sum_{p \in \rm Bins} P(p_{\rm num}|p)}.
\end{equation}
To extract the extent of the error bars below and above $p_{\rm num}$ we determine the $p$ values at which the distribution $P(p|p_{\rm num})$ crosses the value $1/\sqrt{e}$. The error bars determined by this approach coincide with the standard deviation $\sigma$ calculated from Eq.~(\ref{Eq:BinomialSTD}) in that equation's range of validity.

\section{Results}
\label{Sec:Results}

We now present the results of our numerical calculations. We analyze the statistical properties of randomly sampled quantum gate sequences, and we demonstrate how these results can be used to establish a probabilistic approach to quantum circuit synthesis.

\subsection{Establishing multiplicity of equivalent quantum circuits for small systems}

As a starting point for the present study, we consider the small system sizes treated in Ref.~\cite{AshhabQuantumCircuits}. In most of the figures in this manuscript, we plot the probability ($p$) of unit-fidelity quantum circuits as a function of $N$. In each plot, the $x$ axis starts from the relevant lower bound, i.e.~Eq.~(\ref{Eq:LowerBoundStatePrep}), (\ref{Eq:LowerBoundUnitary}), (\ref{Eq:LowerBoundStatePrepB}) or (\ref{Eq:LowerBoundUnitaryB}).

\begin{figure}[h]
\includegraphics[width=7.50cm]{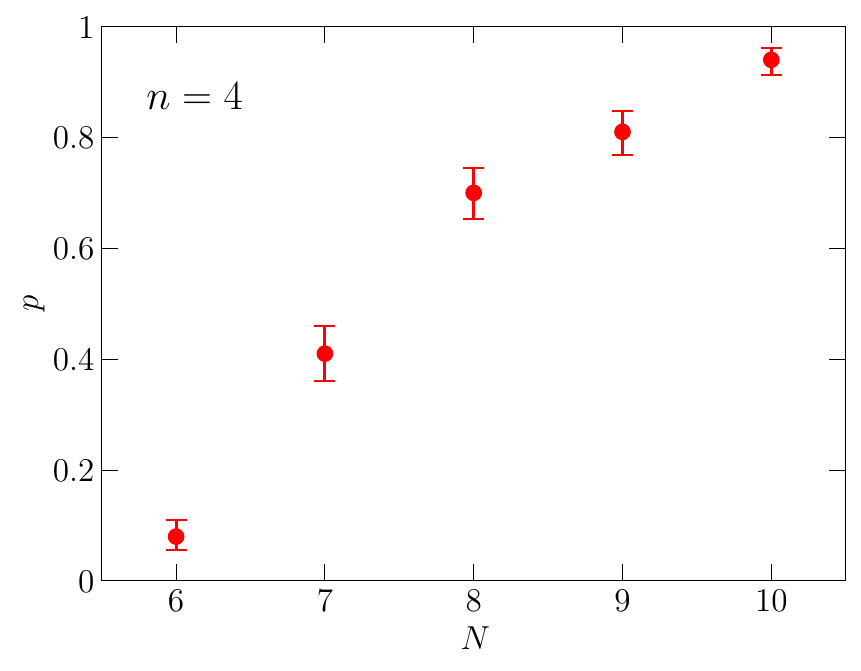}
\includegraphics[width=7.50cm]{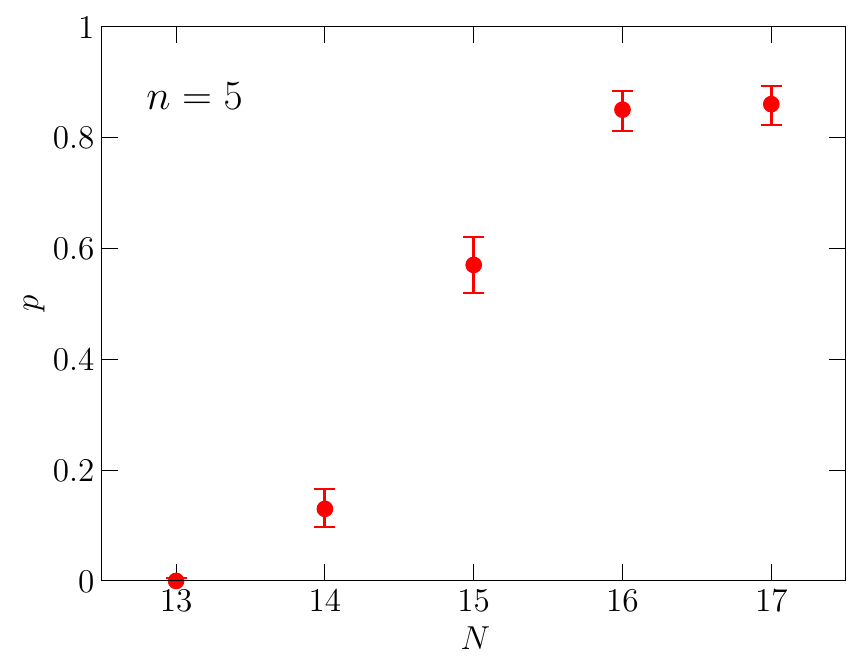}
\includegraphics[width=7.50cm]{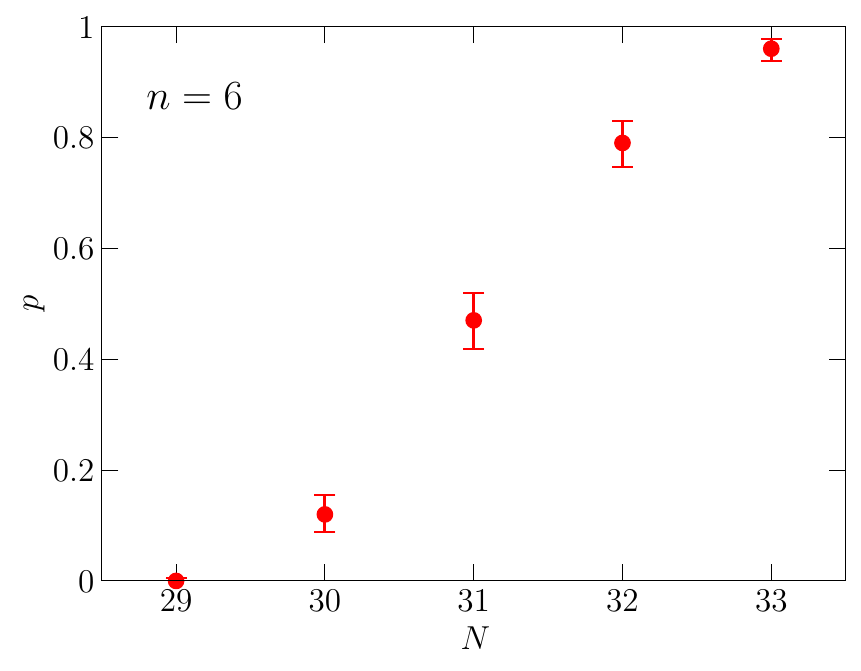}
\includegraphics[width=7.50cm]{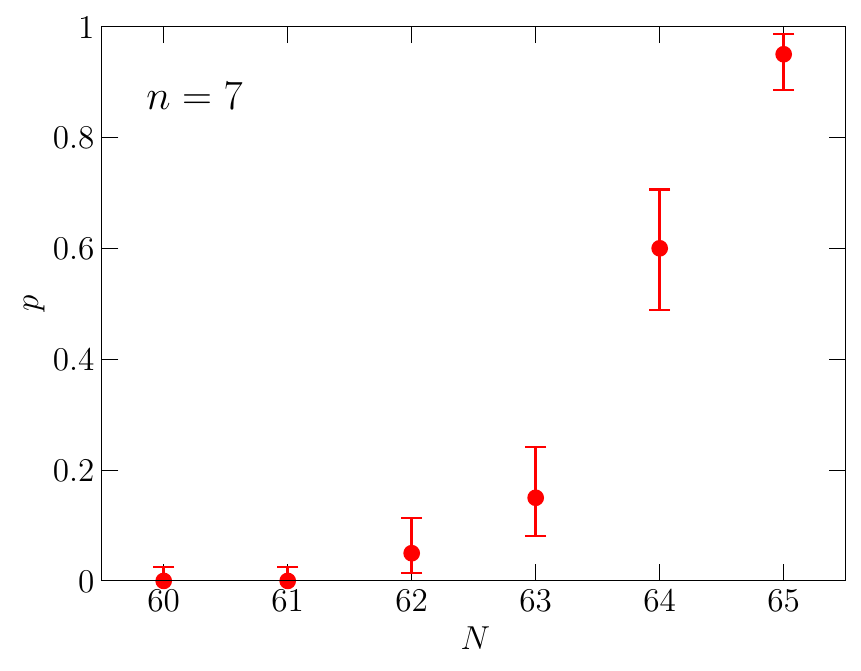}
\includegraphics[width=7.50cm]{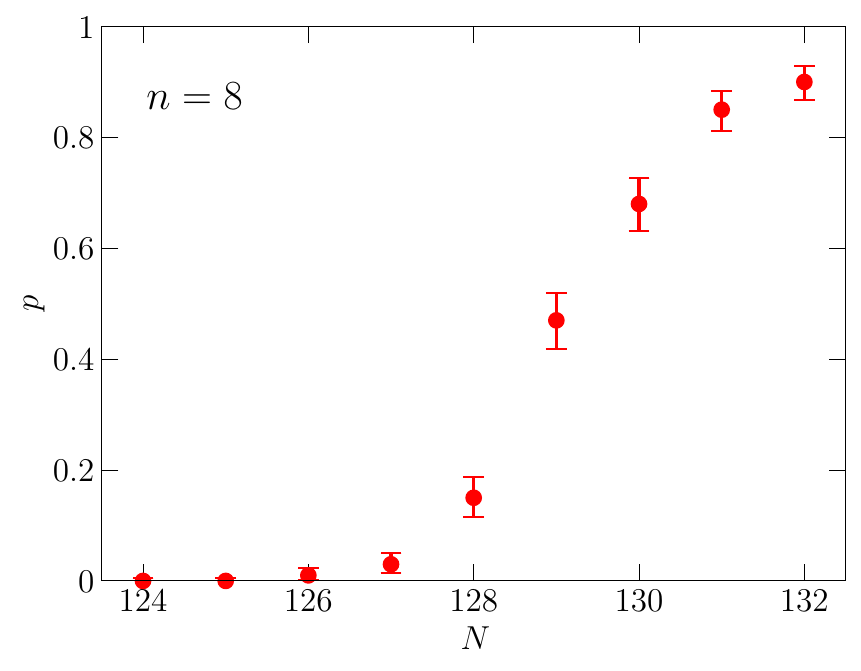}
\caption{Probability $p$ that a random quantum circuit gives $F=1$ as a function of circuit size $N$ for the case of state preparation of an $n$-qubit system using the CNOT gate as the entangling gate. The $F=1$ probability rises rapidly from zero to almost 1 even when $N$ corresponds to just a few gates more than the theoretical lower bound $N_{\rm LB, SP, CNOT}$. The error bars account for the uncertainty in inferring the probability $p$ from the fraction $p_{\rm num}$ obtained in the random sampling calculation.}
\label{Fig:FractionStateCNOT}
\end{figure}

\begin{figure}[h]
\includegraphics[width=7.5cm]{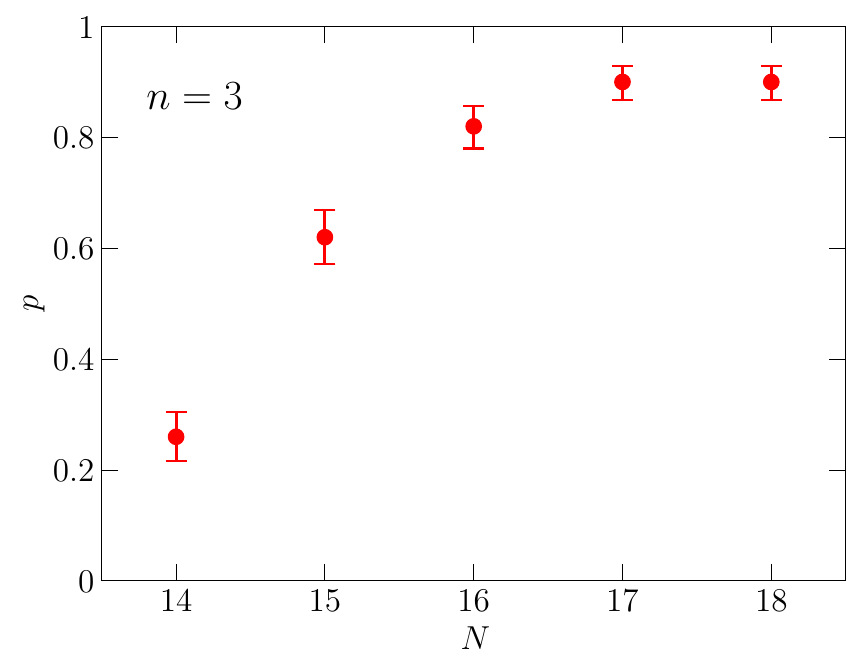}
\includegraphics[width=7.5cm]{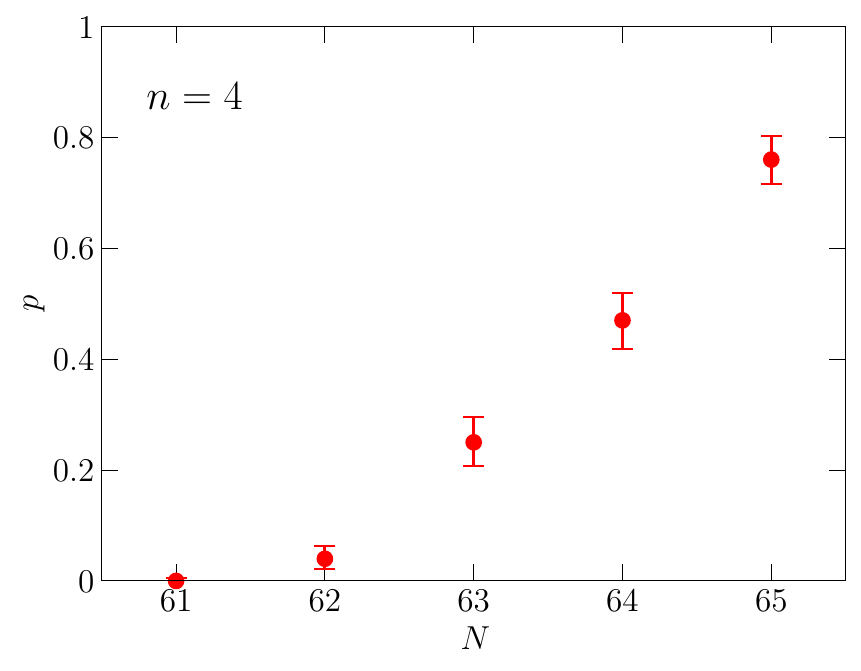}
\includegraphics[width=7.5cm]{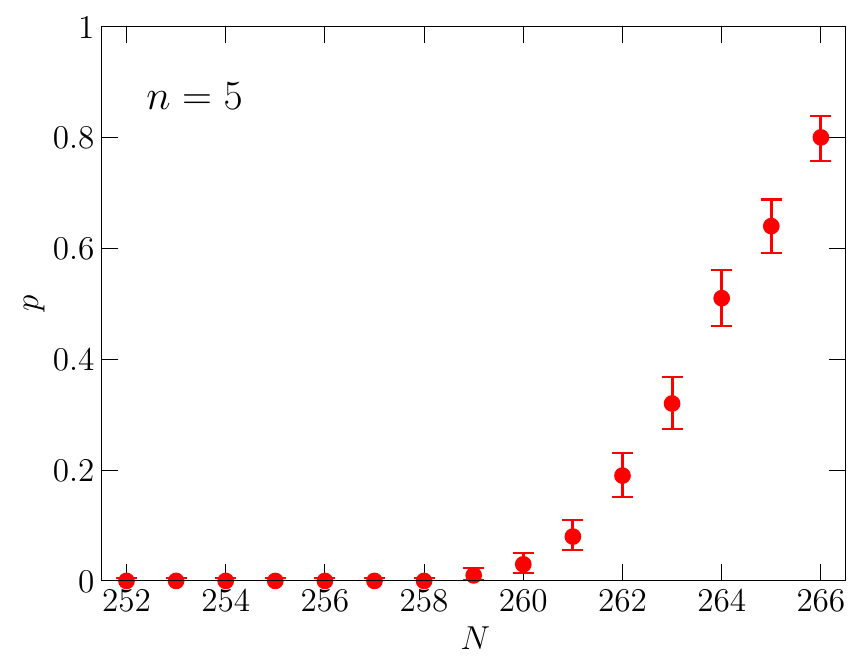}
\caption{Same as in Fig.~\ref{Fig:FractionStateCNOT}, but for the case of unitary operator synthesis.}
\label{Fig:FractionUnitaryCNOT}
\end{figure}

The results for the case of state preparation are shown in Fig.~\ref{Fig:FractionStateCNOT}. For 4 qubits ($n=4$), the minimum number of CNOT gates needed to achieve $F=1$ is $N=6$. Even at the minimum required value of $N$, i.e.~$N=6$, 8\% of the quantum circuits have $F=1$. If we increase $N$, $p$ increases rapidly, such that at $N=10$, we find that 94\% of the quantum circuits have $F=1$. In other words, if we randomly generate a quantum circuit with $n=4$ and $N=10$ and optimize the single-qubit rotation parameters, the quantum circuit will almost certainly give $F=1$. In the unlikely case that the first attempt does not succeed, a second random quantum circuit can be generated. The probability of finding at least one $F=1$ quantum circuit approaches unity very rapidly with increasing number of trial quantum circuits. Specifically, the success probability after $N_{\rm trial}$ is given by
\begin{equation}
P_{\rm success} = 1-(1-p)^{N_{\rm trial}},
\label{Eq:SuccessProbability}
\end{equation}
such that the failure probability decreases exponentially with $N_{\rm trial}$.

We obtain similar results for the case of unitary operator synthesis, as shown in Fig.~\ref{Fig:FractionUnitaryCNOT}. For $n=3$, the minimum number of CNOT gates needed to achieve $F=1$ is $N=14$. The $F=1$ probability in Fig.~\ref{Fig:FractionUnitaryCNOT} ($n=3$) rises from 26\% at $N=14$ to 90\% at $N=17$. As in the case of state preparation, the fraction of quantum circuits that give $F=1$ increases rapidly as soon as the number of CNOT gates exceeds the minimum required number.

\subsection{Larger systems with the CNOT gate as the entangling gate}

For state preparation with $n=5$, the parameter-counting lower bound in Eq.~(\ref{Eq:LowerBoundStatePrep}) is $N_{\rm LB, SP, CNOT}=13$. However, none of the 100 random trial quantum circuits that we tried gave $F=1$. For $N=14$, 13\% of the quantum circuits gave $F=1$ and the fraction increased rapidly to 86\% at $N=17$. This result suggests that the minimum number of CNOT gates needed for perfect state preparation is $N=14$, slightly higher than the value $N_{\rm LB, SP, CNOT}=13$ given by Eq.~(\ref{Eq:LowerBoundStatePrep}). Similarly, for $n=6$, our numerical calculations suggest that the minimum number of CNOT gates is higher than that of Eq.~(\ref{Eq:LowerBoundStatePrep}) by one CNOT gate. For $n=7$, we find a minimum number of CNOT gates that is higher than that of Eq.~(\ref{Eq:LowerBoundStatePrep}) by two CNOT gates. We note here that, because of the significant computational cost (as described in Table \ref{Table:ResourceEstimates}), we used only 20 random quantum circuits for each parameter set in the case of $n=7$. For $n=8$, we also obtain a minimum number ($N=126$) that is higher than that of Eq.~(\ref{Eq:LowerBoundStatePrep}) by two CNOT gates. In all cases shown in Fig.~\ref{Fig:FractionStateCNOT}, the fraction of $F=1$ quantum circuits increases rapidly from zero to near one. This result gives us confidence that the premise of this work, i.e.~that there is large multiplicity in the unit-fidelity quantum circuits even at or close to the minimum required circuit size, is valid. Although Eq.~(\ref{Eq:LowerBoundStatePrep}) is consistently not tight for $n\geq 5$, it provides a very good estimate even when $N\sim 100$. The deviation between the formula and our numerically obtained values of $N$ grows much more slowly than $N$.

\begin{figure}[h]
\includegraphics[width=7.5cm]{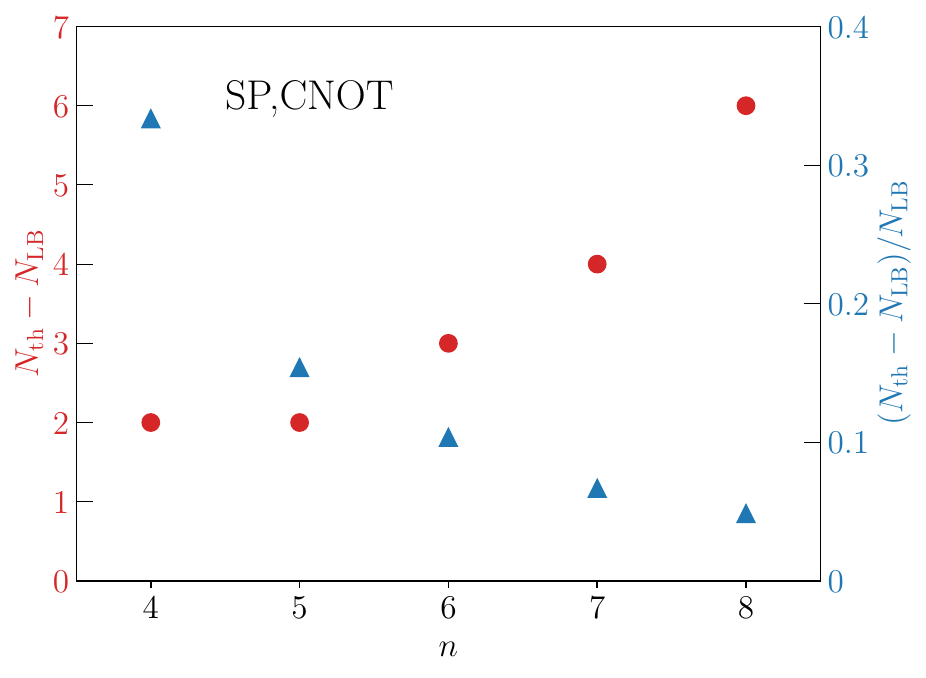}
\includegraphics[width=7.5cm]{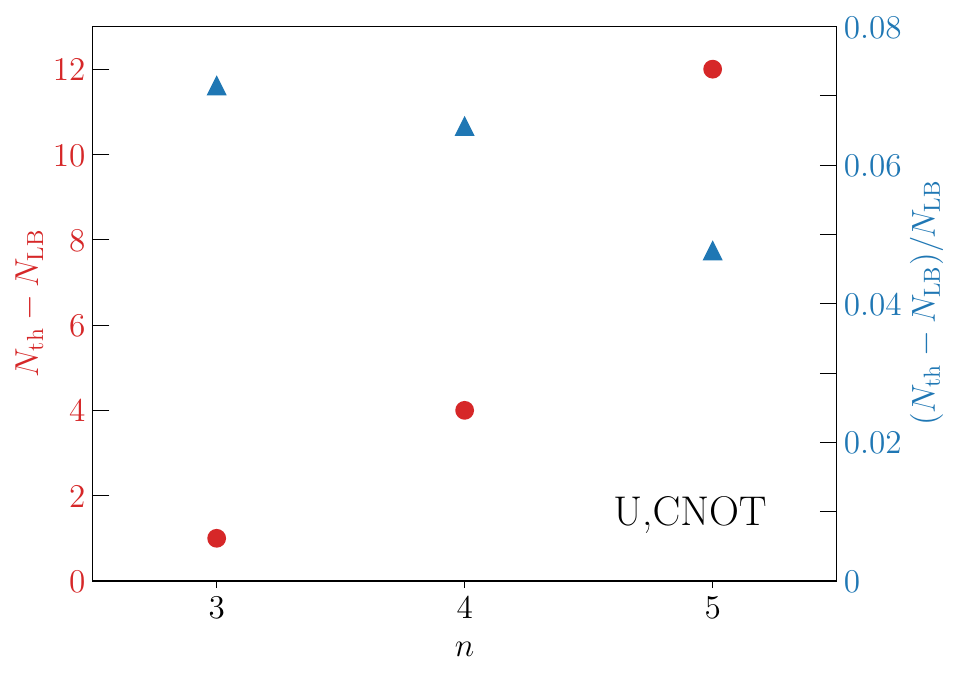}
\includegraphics[width=7.5cm]{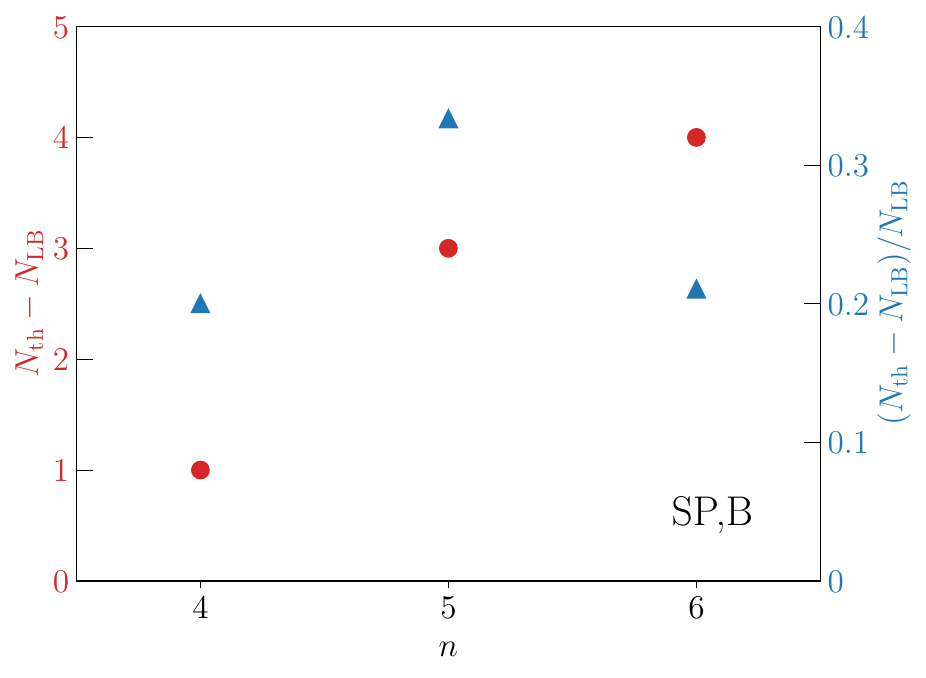}
\includegraphics[width=7.5cm]{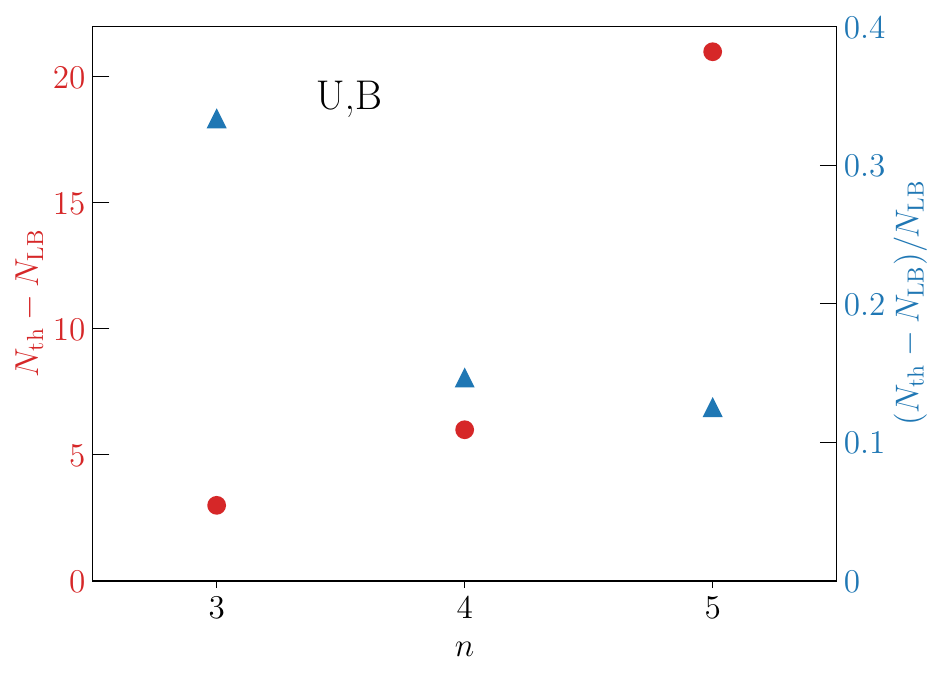}
\caption{Threshold at which the success probability exceeds 50\%. The different panels are identified by their legends: SP corresponds to state preparation, while U corresponds to unitary operator synthesis. Similarly, CNOT corresponds to using the CNOT gate, while B corresponds to using the $B$ gate, which is discussed in Sec.~\ref{Sec:Results}.C. The red circles show $N_{\rm th}-N_{\rm LB}$ ($y$ axis on the left-hand side), while the blue triangles show $(N_{\rm th}-N_{\rm LB})/N_{\rm LB}$ ($y$ axis on the right-hand side). In almost all cases, $(N_{\rm th}-N_{\rm LB})/N_{\rm LB}$ decreases with increasing $n$, which means that the ratio $N_{\rm th}/N_{\rm LB}$ gradually approaches 1, which means that in relative terms $N_{\rm th}$ approaches $N_{\rm LB}$ with increasing system size.}
\label{Fig:Threshold}
\end{figure}

As alluded to in the previous subsection, in addition to the question of identifying the smallest value of $N$ at which we obtain at least one quantum circuit with $F=1$, another important question is the value of $N$ at which the probability $p$ becomes on the order of 1. In this context, we define the threshold number $N_{\rm th}$ as the smallest value of $N$ at which $p>50$\%. The value of $N_{\rm th}$ can be inferred from a plot of $p$ as a function of $N$. The results for the cases of Figs.~\ref{Fig:FractionStateCNOT} and \ref{Fig:FractionUnitaryCNOT} are shown in Fig.~\ref{Fig:Threshold}. Equation (\ref{Eq:SuccessProbability}) indicates that above $N_{\rm th}$ it should be easy to find a unit-fidelity quantum circuit with just a few attempts, since the success probability in each attempt is above 50\%.

For the case of unitary operator synthesis (Fig.~\ref{Fig:FractionUnitaryCNOT}), for $n=4$, the $F=1$ fraction is zero at the lower bound $N_{\rm LB, U, CNOT}=61$. However, it is 4\% at $N=62$ and rises rapidly to reach 76\% at $N=65$. For $n=5$, the lower bound $N_{\rm LB, U, CNOT}=252$. However, we obtain the first $F=1$ quantum circuit at $N=259$, i.e.~seven more CNOT gates than $N_{\rm LB, U, CNOT}$. Considering that $N_{\rm LB, U, CNOT}=252$, seven gates is under 3\% of $N_{\rm LB, U, CNOT}$. This result strongly suggests that the lower bound is not tight in this case. It is also remarkable that the probability $p$ rises to 80\% at $N=266$, such that $N_{\rm th}$ is only a few steps above the smallest value of $N$ that gives $F=1$, even though $N\approx 260$.

It should be noted that there is a possibility that larger values of $n$ lead to slower convergence, which could in turn lead to lowering the $F=1$ fraction and an overestimation of the minimum number of gates needed to achieve $F=1$. Our past experience with these optimization calculations does not indicate that such a slow convergence effect occurs. In fact, the increased number of control parameters can speed up the convergence. Nevertheless, as is the case with many search algorithms, we cannot completely rule out this possibility. It should also be noted that the optimization algorithm occasionally gives somewhat lower values, e.g.~$F=0.999$, for quantum circuits that correspond to $F=1$. This property was observed in Ref.~\cite{AshhabQuantumCircuits}; it manifests itself when the algorithm is run multiple times for the same CNOT gate configuration. However, this effect should only reduce the $F=1$ fraction by a finite factor and should therefore not affect our conclusions about the minimum value of $N$ that allows achieving $F=1$.

One might also wonder if it is possible that for large quantum circuit sizes, e.g.~in the case of unitary operator synthesis with $n=5$ and $N\approx 260$, it might be possible to obtain fidelities higher than our threshold of $1-10^{-8}$ even if the gate configuration in question does not allow achieving $F=1$. In other words, 260 CNOT gate along with the single-qubit gates in the circuit allow so much freedom that one can get a very good approximation for the target. We do not expect that this possibility is affecting our calculations, because in our results the threshold is exceeded only for $N$ values higher than the theoretical lower bounds in Eqs.~(\ref{Eq:LowerBoundStatePrep}) and (\ref{Eq:LowerBoundUnitary}). In other words, if false positives occur for $N\approx 260$, we would expect them to occur for $N\approx 250$ as well, considering the small relative difference between $N=250$ and $N=260$. We have not seen any such false positives below the lower bounds for $N$.

\subsection{Using the $B$ gate as the entangling gate}

\begin{figure}[h]
\includegraphics[width=8.0cm]{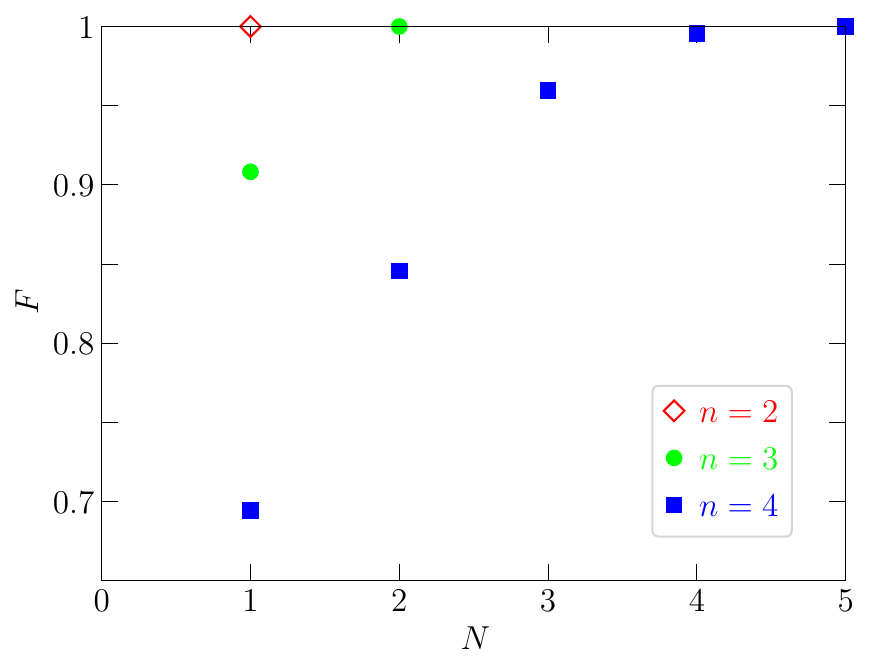}
\caption{Maximum achievable fidelity $F$ as a function of the number of $B$ gates for state preparation with $n=2$ (red diamond), 3 (green circles) and 4 (blue squares). By comparing these results with those of Ref.~\cite{AshhabQuantumCircuits}, we can see that the $B$ gate allows achieving the same results as the CNOT gate with smaller values of $N$. In the case of the CNOT gate, $F=1$ is achieved for $N=1$, 3 and 6 instead of 1, 2 and 5, respectively.}
\label{Fig:FidelityStatePrepB}
\end{figure}

\begin{figure}[h]
\includegraphics[width=8.0cm]{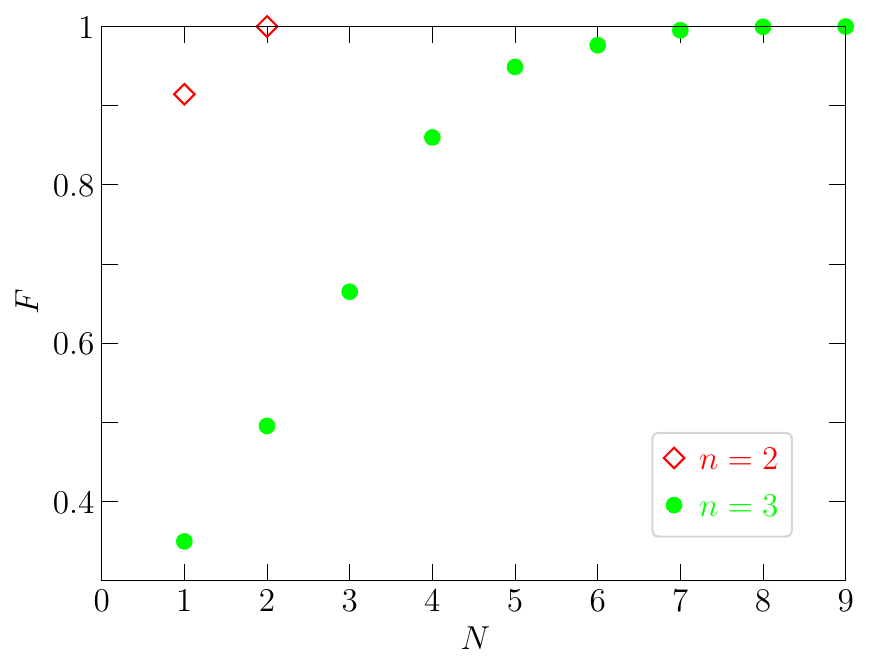}
\caption{Maximum achievable fidelity $F$ as a function of the number of $B$ gates for unitary operator synthesis with $n=2$ (red diamonds) and 3 (green circles). In the case of the CNOT gate analyzed in Ref.~\cite{AshhabQuantumCircuits}, $F=1$ is achieved for $N=3$ and 14 instead of 2 and 9, respectively.}
\label{Fig:FidelityUnitaryB}
\end{figure}

\begin{figure}[h]
\includegraphics[width=7.5cm]{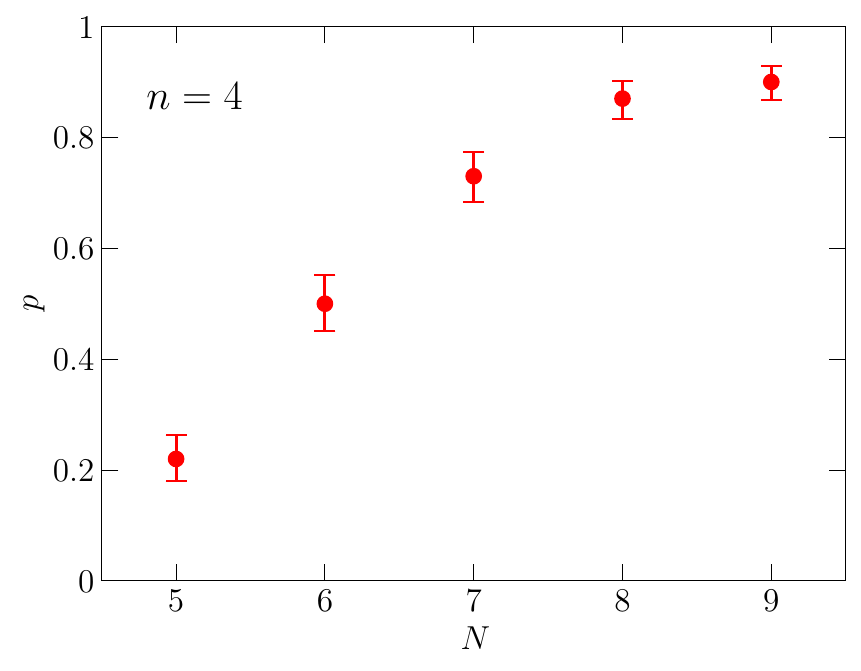}
\includegraphics[width=7.5cm]{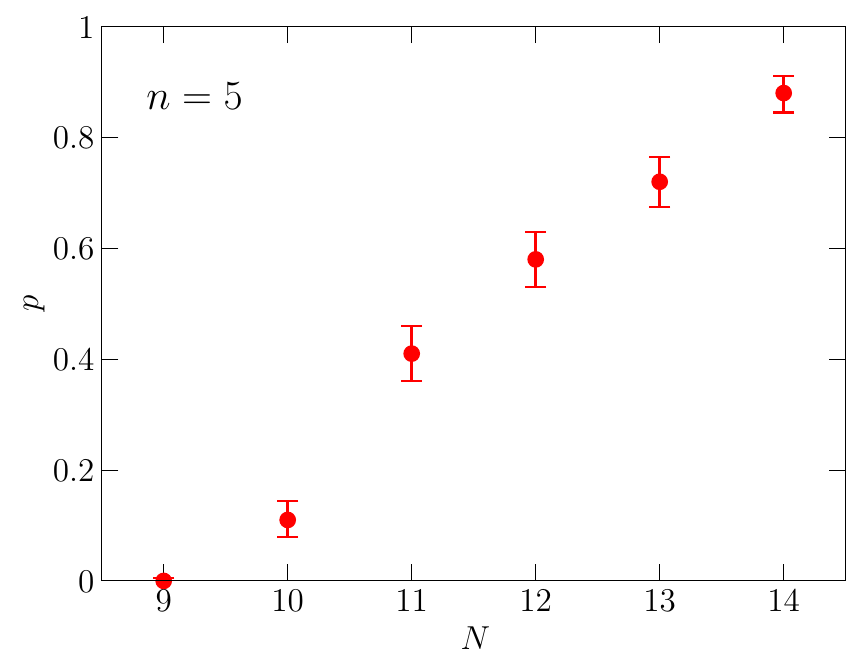}
\includegraphics[width=7.5cm]{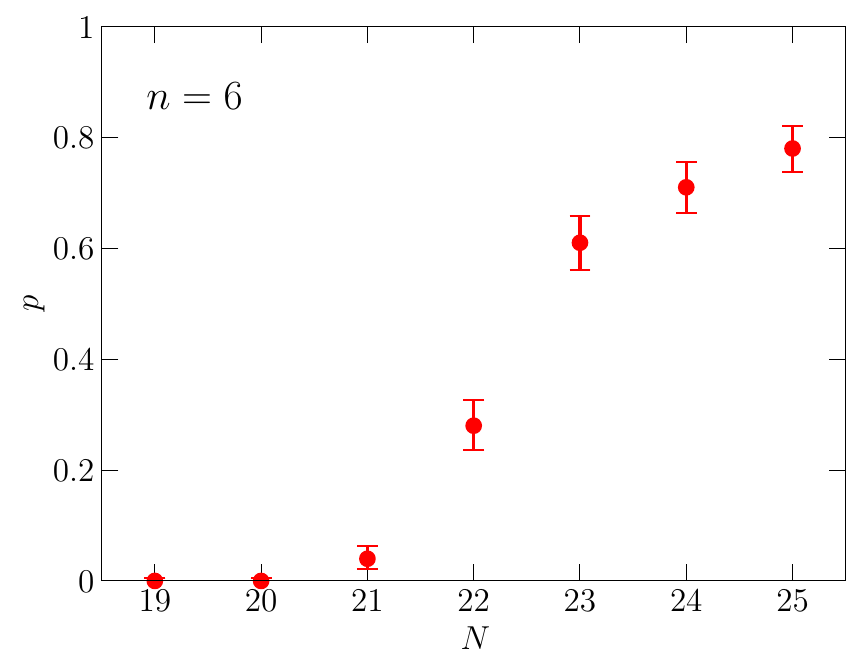}
\caption{Same as in Fig.~\ref{Fig:FractionStateCNOT}, but for the case where the $B$ gate is used as the entangling gate. The $B$ gate allows achieving the same results as the CNOT gate with smaller values of $N$.}
\label{Fig:FractionStateB}
\end{figure}

\begin{figure}[h]
\includegraphics[width=7.5cm]{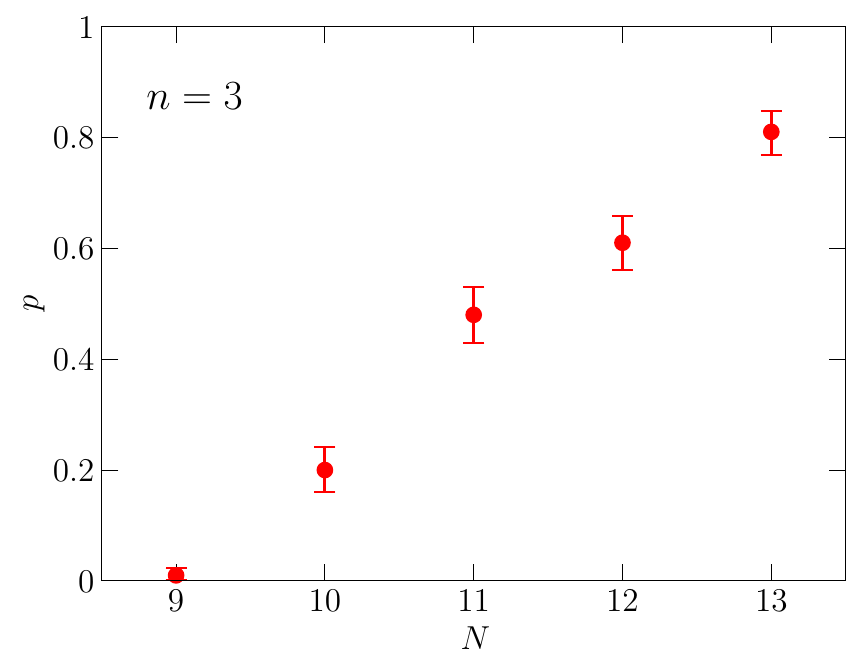}
\includegraphics[width=7.5cm]{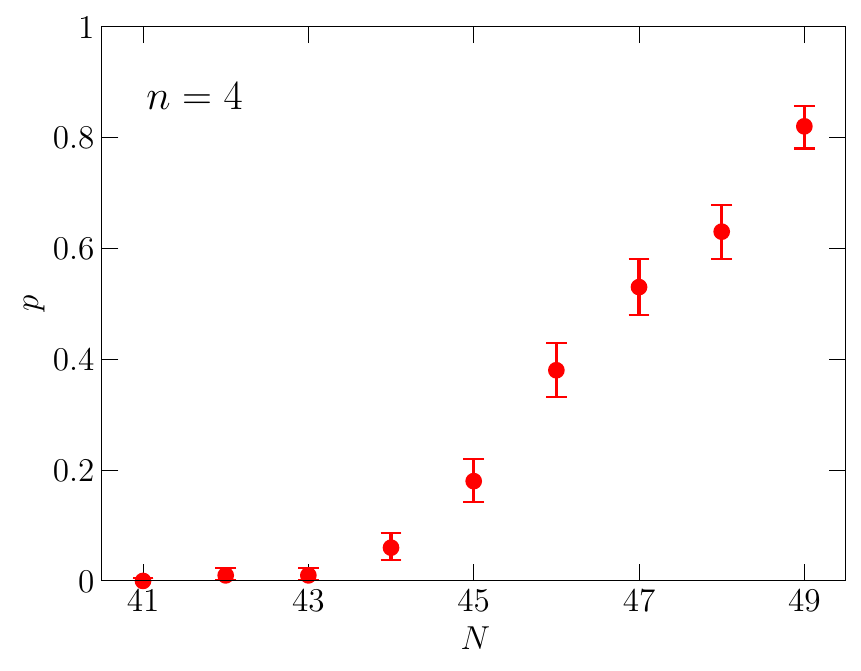}
\includegraphics[width=7.5cm]{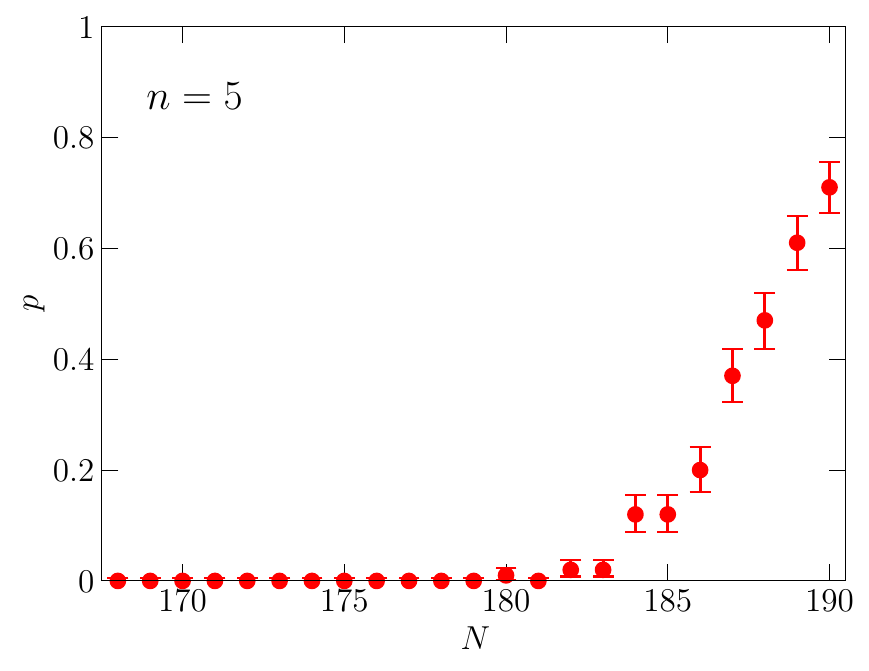}
\caption{Same as in Fig.~\ref{Fig:FractionUnitaryCNOT}, but for the case where the $B$ gate is used as the entangling gate. The $B$ gate allows achieving the same results as the CNOT gate with smaller values of $N$.}
\label{Fig:FractionUnitaryB}
\end{figure}

Although the CNOT gate is the standard entangling gate used in quantum algorithm design, the so-called $B$ gate \cite{ZhangBGate}

\

\centerline{
\Qcircuit @C=1em @R=.7em {
& \multigate{1}{B} & \qw & & & \ctrl{1} & \gate{e^{i\frac{\pi}{4}\sigma_x}} & \qw \\
& \ghost{B}& \qw & \ {}^{^{^{^{^{^{^{^{\displaystyle \sim}}}}}}}} & & \targ & \ctrl{-1} & \qw \\
}
}

\

\noindent has one important advantage over the CNOT gate for the purpose of our study. Using the $B$ gate can lead to smaller quantum circuit sizes to achieve the same result as using the CNOT gate. This point can be understood by considering the parameter-counting calculation used to determine the lower bounds for $N$. The CPhase gate, which is equivalent to the CNOT gate up to single-qubit rotations, commutes with single-qubit rotations about the $z$ axis. This property means that all $z$-axis rotations acting on the same qubit can be merged into one rotation. As a result, the number of independent quantum circuit parameters is reduced. If we use an entangling gate that does not commute with any single-qubit rotations, we maximize the number of free adjustable parameters in a quantum circuit of size $N$, thus increasing the number of free parameters per single-qubit rotation from 2 to 3 (excluding the first layer of single-qubit rotations). The $B$ gate does not commute with any single-qubit rotations. The parameter counting calculation then gives the lower bounds
\begin{equation}
N_{{\rm LB, SP}, B} = \left\lceil \frac{2^{n}-1-n}{3} \right\rceil
\label{Eq:LowerBoundStatePrepB}
\end{equation}
for state preparation and
\begin{equation}
N_{{\rm LB, U}, B} = \left\lceil \frac{4^n-1-3n}{6} \right\rceil
\label{Eq:LowerBoundUnitaryB}
\end{equation}
for unitary operator synthesis. It is indeed well-known that the $B$ gate allows the synthesis of arbitrary two-qubit gates more efficiently than the CNOT gate: two $B$ gates compared to 3 CNOT gates. With this point in mind, we perform similar calculations to those presented above but using the $B$ gate as the entangling gate, along with the set of all single-qubit rotations.

As in the case of the CNOT gate, we can use the brute-force approach in which we analyze all possible gate configurations for state preparation up to $n=4$ and for unitary operator synthesis up to $n=3$. The results are shown in Figs.~\ref{Fig:FidelityStatePrepB} and \ref{Fig:FidelityUnitaryB}. These figures show that the $B$ gate is more efficient than the CNOT gate in the sense that it allows quantum circuits with smaller values of $N$ to perform the same operations. In the case of state preparation, the minimum numbers of CNOT gates needed for $n=2$, 3 and 4 are 1, 3 and 6, respectively. Using the $B$ gate, we obtain the values 1, 2 and 5, respectively. The theoretical lower bounds are $N_{{\rm LB, SP}, B}=1$, 2 and 4, respectively. In other words, $N_{{\rm LB, SP}, B}$ in Eq.~(\ref{Eq:LowerBoundStatePrepB}) is achieved for $n=2$ and 3 but not for $n=4$. In the case of unitary operator synthesis, the minimum numbers of CNOT gates needed for $n=2$ and 3 are 3 and 14, respectively. Using the $B$ gate, the numbers are 2 and 9, respectively, both matching the theoretical lower bounds given by Eq.~(\ref{Eq:LowerBoundUnitaryB}).

The results of the probabilistic search calculations using the $B$ gate are shown in Figs.~\ref{Fig:FractionStateB} and \ref{Fig:FractionUnitaryB}. In all cases, the $x$ axis starts at the lower bound given by Eq.~ (\ref{Eq:LowerBoundStatePrepB}) or (\ref{Eq:LowerBoundUnitaryB}). The results are generally similar to those obtained using the CNOT gate. The probability $p$ rises above zero either at the lower bound value or slightly above it, and $p$ reaches close to one after just a few additional steps, i.e.~for a slightly higher value of $N$. Here it is worth taking a closer look at the cases $n=4,N=42,43$ and $n=5,N=180$. Considering that only one quantum circuit gave $F=1$ in each one of the cases, one might wonder if these were cases in which the fidelity was slightly higher than $1-10^{-8}$, giving false positive counts of perfect quantum circuits. However, in two out of these three cases $1-F$ was below $10^{-12}$. In the case $n=4,N=42$, the quantum circuit identified as having $F=1$ in fact had $1-F\sim 10^{-9}$. However, we reran the optimization algorithm for this quantum circuit five times, and two of these reruns gave $1-F < 10^{-12}$, which strongly suggests that these cases were not false positives. These results suggest that the fraction of quantum circuits that give $F=1$ at the minimum required number of entangling gates decreases with increasing system size. This behavior differs from our observation for small systems, where a significant fraction of quantum circuits gave $F=1$ even at the theoretical lower bound for circuit size. Nevertheless, our probabilistic approach remains a good method to find quantum circuits with small numbers of gates and high fidelities.

\subsection{Practical considerations}

One important question when comparing different entangling gates, e.g.~the CNOT and $B$ gate, is the ability to perform the entangling gate in a real setup. The CNOT gate between two neighboring qubits can be implemented in realistic systems with a simple control pulse \cite{Paraoanu,Rigetti,DeGroot}. Even if we include technical difficulties such as coupling to nearby qubits and weak anharmonicity, one can still implement the CNOT gate with relatively simple pulses, as explained in Refs.~\cite{DeGrootTheory,AshhabSpeedLimits2022}. To assess the ease of implementing the $B$ gate, we first express it in matrix form as
\begin{equation}
B = \left(
\begin{array}{cccc}
1 & 0 & 0 & 0 \\
0 & \frac{1}{\sqrt{2}} & \frac{i}{\sqrt{2}} & 0 \\
0 & 0 & 0 & 1 \\
0 & \frac{i}{\sqrt{2}} & \frac{1}{\sqrt{2}} & 0
\end{array}
\right),
\end{equation}
where the rows and columns are ordered as in the computational basis $\left\{ \ket{00}, \ket{01}, \ket{10}, \ket{11} \right\}$. If we would like to implement this operation in a single step using Hamiltonian evolution of the form $e^{-iHt}$, where $H$ is the Hamiltonian and $t$ is the evolution time, the Hamiltonian will need to have the form
\begin{equation}
H \propto i \log B =
\left(
\begin{array}{cccc}
0 & 0 & 0 & 0 \\
0 & -0.248 & -0.372 + 0.599 i & -0.372 - 0.599 i \\
0 & -0.372 - 0.599 i & -1.447 & 1.447 + 0.154 i \\
0 & -0.372 + 0.599 i &  1.447 - 0.154  i & -1.447
\end{array}
\right).
\end{equation}
This Hamiltonian describes couplings with very specific strength ratios between all pairings of the states $\ket{01}$, $\ket{10}$ and $\ket{11}$. There is no simple physical mechanism that corresponds to this Hamiltonian and naturally implements the $B$ gate. As a result, the $B$ gate would likely have to be implemented in a two-step process as illustrated in the diagram at the beginning of Sec.~\ref{Sec:Results}.C. If we perform the gate count with each $B$ gate counting as two controlled operations, and we compare the resulting numbers to the corresponding numbers for the CNOT gate, we find that using the CNOT gate is more efficient in practice. Furthermore, in a real setup, the naturally occurring interactions play a crucial role in determining the most natural entangling gate. Any other entangling gate can then be constructed from that platform-specific natural entangling gate.

Another important point relates to finding quantum circuits for different targets. As demonstrated in Ref.~\cite{AshhabQuantumCircuits}, if a CNOT-gate-based quantum circuit allows perfect implementation of one arbitrary target, i.e.~a target that is not coincidentally easy to prepare, the same CNOT gate configuration allows perfect implementation of any other target. Only the single-qubit rotation parameters need to be calculated for each new target. In other words, once we identify one unit-fidelity quantum circuit via the random combinatorial search technique, we do not need to run the same random search again for the same system size.

Another point relates to the high fidelity threshold in our calculations. In a practical setting, especially in present-day or near-future applications, there might not be any need to design extremely high-fidelity operations. Indeed, realistic decoherence considerations favor short quantum circuit. One might therefore find it advantageous to use a quantum circuit that has a theoretical fidelity below 1, if this quantum circuit is substantially shorter than the $F=1$ quantum circuit. Our probabilistic approach can be used in this case as well. It is also possible to include decoherence in the GRAPE algorithm \cite{Khaneja}. In this case, we can expect to obtain a maximum in $F$ as a function of $N$. When $N$ becomes too large, decoherence disrupts the dynamics and results in low fidelities.

\subsection{Constructing the $n$-qubit Toffoli gate}

\begin{figure}[h]
\includegraphics[width=8.0cm]{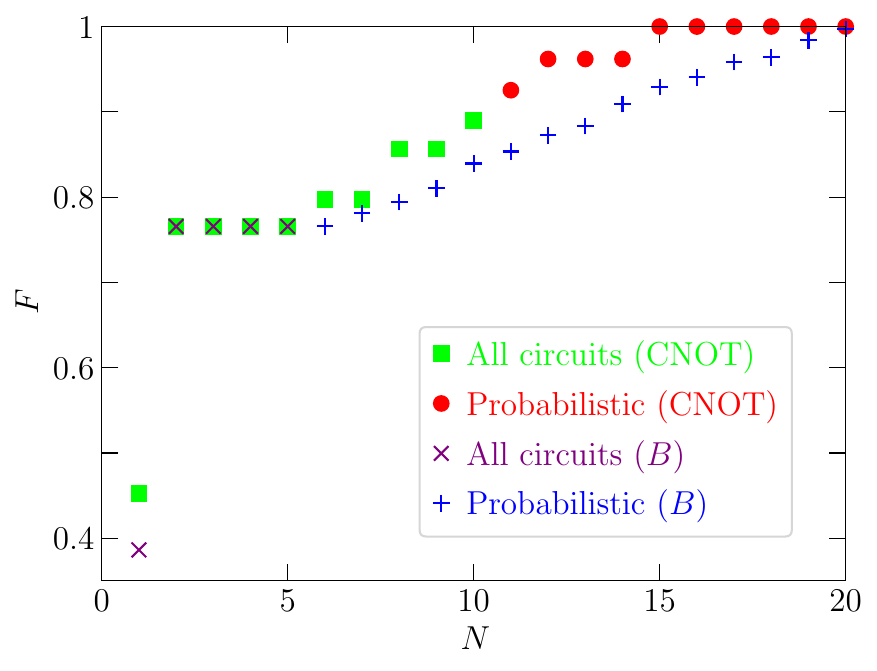}
\caption{Maximum achievable fidelity $F$ as a function of the number of entangling (CNOT or $B$) gates used to synthesize a 4-qubit Toffoli gate. The green squares correspond to the results obtained in Ref.~\cite{AshhabQuantumCircuits}, where all possible CNOT gate configurations were analyzed. The purple $\times$ symbols are obtained by analyzing all possible $B$ gate configurations. The red circles and blue $+$ symbols correspond to the random search approach for the CNOT gate and $B$ gate, respectively. The 4-qubit Toffoli gate can be perfectly synthesized using 15 CNOT gates. Unlike what we found for the case of arbitrary $n$-qubit gate synthesis, the $B$ gate seems to be less efficient than the CNOT gate in this case. At $N=20$, the $B$ gate gives $F=0.997$.}
\label{Fig:FidelityToffoli}
\end{figure}

In the previous subsections we considered the case of arbitrary target operations. As such, the target states and unitary operators were generated using random number generators. In this subsection, we consider synthesizing multi-qubit Toffoli gates. Specifically, we search for quantum circuits that implement the 4-qubit Toffoli gate using 2-qubit entangling gates. The lower bounds according to Eqs.~(\ref{Eq:LowerBoundUnitary}) and (\ref{Eq:LowerBoundUnitaryB}) are $N_{\rm LB, U, CNOT}=61$ and $N_{{\rm LB, U}, B}=41$. However, considering that the $n$-qubit Toffoli gate has a special and simple structure, we can expect that it can be obtained using quantum circuits of smaller sizes. The results of our numerical calculations are shown in Fig.~\ref{Fig:FidelityToffoli}. The green squares show the results of Ref.~\cite{AshhabQuantumCircuits}, which were limited to $N\leq 10$ because larger values of $N$ would have taken many years of single-core computation time. Using the random search approach (with $10^4$ random quantum circuits for each set of parameters), we find that the 4-qubit Toffoli gate can be decomposed into 15 CNOT gates. We note here that the $F=1$ fraction was very low: for $N=15,16,\dots,20$, we found, respectively, 1, 2, 7, 31, 56 and 133 perfect-fidelity circuits out of $10^4$ random circuits in each set. The low $F=1$ fraction in this case contrasts with the situation for the 3-qubit Toffoli gate. In Ref.~\cite{AshhabQuantumCircuits}, we found that about 7\% of all gate configurations with $N=6$ gave $F=1$. The CNOT gate configuration in the quantum circuit that gave $F=1$ for $N=15$ is described by the diagram:

\centerline{
\Qcircuit @C=1em @R=.7em {
& \qw & \ctrl{2} & \qw & \ctrl{3} & \qw & \qw & \ctrl{2} & \ctrl{1} & \qw & \ctrl{2} & \qw & \ctrl{3} & \qw & \ctrl{3} & \ctrl{1} & \qw \\
& \ctrl{1} & \qw & \qw & \qw & \ctrl{2} & \ctrl{1} & \qw & \ctrl{0} & \ctrl{2} & \qw & \ctrl{1} & \qw & \qw & \qw & \ctrl{0} & \qw \\
& \ctrl{0} & \ctrl{0} & \ctrl{1} & \qw & \qw & \ctrl{0} & \ctrl{0} & \qw & \qw & \ctrl{0} & \ctrl{0} & \qw & \ctrl{1} & \qw & \qw & \qw \\
& \qw & \qw & \ctrl{0} & \ctrl{0} & \ctrl{0} & \qw & \qw & \qw & \ctrl{0} & \qw & \qw & \ctrl{0} & \ctrl{0} & \ctrl{0} & \qw & \qw
}
}

\

\noindent where single-qubit rotations are not included in the diagram. This circuit does not have any easily discernible pattern. The circuit is slightly larger than the one in Ref.~\cite{Schuch}, which gave a decomposition of the 4-qubit Toffoli gate into 14 CNOT gates, along with single-qubit gates. Hence, the random search approach produced an almost optimal decomposition of the gate. Besides, while the quantum circuit of Ref.~\cite{Schuch} was constructed partly using intuition, the random search approach should work even for target gates that do not have a simple structure. We also performed random search calculations using the $B$ gate. In contrast to the case of an arbitrary target unitary operator, the $B$ gate does not outperform the CNOT gate for any value of $N$. On the contrary, it almost consistently gives lower fidelities for the same number of entangling gates. Even at $N=15$, where the CNOT gate gives $F=1$ up to numerical errors, the $B$ gate gives $F=0.93$. At $N=20$, the $B$ gate gives $F=0.997$.

\begin{figure}[h]
\includegraphics[width=8.0cm]{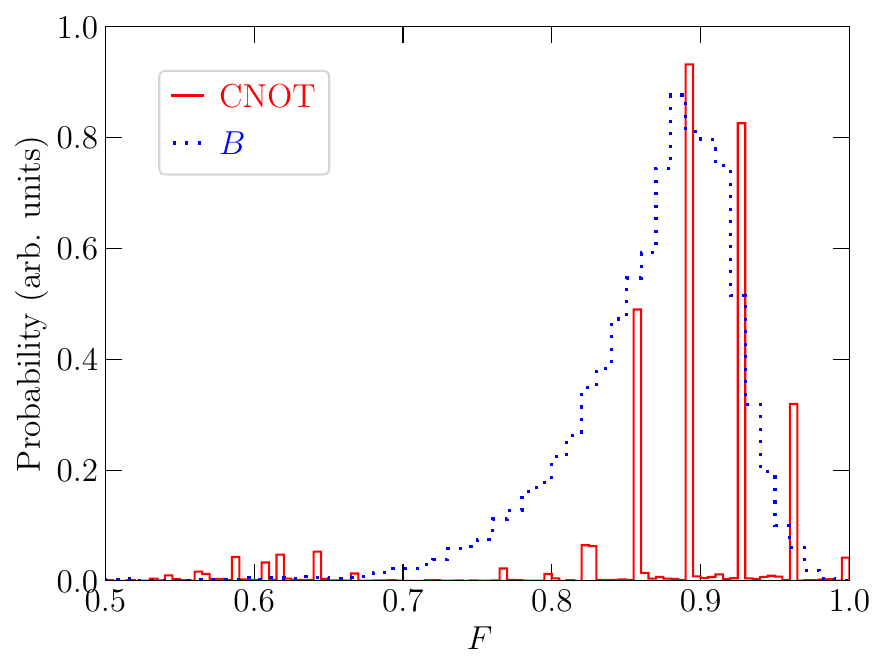}
\caption{Histograms for the fidelity $F$ values for $N=20$ when using the CNOT gate (red solid line) or the $B$ gate (blue dotted line) to synthesize a 4-qubit Toffoli gate. The y-axis scale is not the same for the two curves. For the CNOT gate there are special values of $F$ that dominate the distribution, while for the $B$ gate there is a continuous, broad distribution of $F$ values.}
\label{Fig:ToffoliHistogram}
\end{figure}

Here it is also interesting to inspect the distribution of fidelity values obtained with the CNOT gate and the $B$ gate. The histograms plotted in Fig.~\ref{Fig:ToffoliHistogram} show that the CNOT gate gives a set of sharp peaks. The peak locations, i.e.~the $F$ values at which the peaks occur, seem to be independent of $N$, at least between 1 and 20. Only the heights of the different peaks change with $N$. In contrast, the $B$ gate gives continuous broad distributions of $F$ values. The entire distribution shifts to larger values of $F$ with increasing $N$.

\section{Conclusion}
\label{Sec:Conclusion}

In conclusion, we have demonstrated the possibility of efficient quantum circuit synthesis using a random combinatorial search. This method allows us to analyze multi-qubit systems of sizes that do not allow an analysis of all possible gate configurations. We find that, as long as the quantum circuit size is larger than the theoretical lower bound for perfect implementation of an arbitrary quantum operation, a randomly generated quantum circuit has a high probability of performing an arbitrary operation after the circuit parameters are optimized. It should be noted that the circuit sizes needed to implement arbitrary $n$-qubit operations scale exponentially with $n$. Specific quantum gates can be implemented with smaller circuits. For the case of the $n$-qubit Toffoli gate, the circuit size is expected to scale quadratically with $n$. We applied the random search method to the 4-qubit Toffoli gate and found a perfect fidelity quantum circuit with 15 CNOT gates. In addition to the CNOT gate, we applied our method to the $B$ gate. The $B$ gate consistently showed better performance for arbitrary targets but worse performance for the 4-qubit Toffoli gate. This contrast sheds light on the utility of the $B$ gate as an alternative to the CNOT gate in quantum circuit synthesis. It should of course be kept in mind that the naturally occurring physical interactions determine which gate to use in a real device. Our results demonstrate that the random search approach provides one more powerful tool that can be used to optimize the implementation of quantum information processing tasks.

\section*{Acknowledgment}

We would like to thank Shunsuke Daimon, Norbert Schuch and Naoki Yamamoto for useful discussions. This work was supported by Japan's Ministry of Education, Culture, Sports, Science and Technology (MEXT) Quantum Leap Flagship Program Grant Number JPMXS0120319794, and by the Center of Innovations for Sustainable Quantum AI (Japan Science and Technology Agency [JST], Grant Number JPMJPF2221). The research used computational resources of the supercomputer Fugaku provided by the RIKEN Center for Computational Science.

\end{document}